\documentclass[twocolumn,PLB,superscriptaddress]{revtex4}%
\usepackage{amsfonts}
\usepackage{amsmath}
\usepackage{amssymb}
\usepackage{graphicx}%
\begin{document}
\title{Island rules for the noncommutative black hole}
\author{Yipeng Liu}
\affiliation{School of Mathematics and Physics, China University of Geosciences, Wuhan
430074, China}
\author{Wei Xu}
\affiliation{School of Mathematics and Physics, China University of Geosciences, Wuhan
430074, China}
\author{Baocheng Zhang}
\email{zhangbaocheng@cug.edu.cn}
\affiliation{School of Mathematics and Physics, China University of Geosciences, Wuhan
430074, China}
\begin{abstract}
In the context of noncommutative black holes, we reconsider the island rule
and reproduce the Page curve. Since the radiation entropy of the
noncommutative black hole will eventually diverge in the absence of islands,
it is crucial to include the contribution of islands in the radiation process
to recover the Page curve. When cutoff surface is far from event horizon, we
find that the existence of islands helps to improve the radiation entropy and
avoid its divergence at the late stage, while the radiation entropy for the
commutative black hole with islands is still divergent in this case. When
cutoff surface is close to event horizon, the existence of islands can lead to
the radiation process consistent with the Page curves for both the commutative
and noncommutative black holes, but it is interesting to obtain a
significantly delay Page time for the noncommutative case, allowing us to
avoid considering the contribution of islands for a sufficiently long period
of time.

\end{abstract}
\maketitle


\section{Introduction}

Since the discover of Hawking radiation \cite{Hawking:1974rv,Hawking:1975vcx},
its properties remain largely unknown. Does radiation carry information?
Hawking's initial calculation indicated that the radiation is a thermal
spectrum, carrying no information, thus violating the unitary principle of
quantum mechanics, which is known as \textit{information loss paradox}.
Conversely, if we adhere to the principle of unitary, the entropy of radiation
should follow Page curve \cite{Page:1993wv}.

Currently, most researchers tend to believe that unitary principle should not
be violated, with the most significant evidence coming from the AdS/CFT
correspondence \cite{Maldacena:1997re}. More precisely, according to the
\textquotedblleft Central Dogma\textquotedblright, the black hole and its
surrounding quantum fields can be treated as an ordinary quantum system up to
the cutoff surface. The gravitational influence beyond cutoff surface is
generally considered negligible. However, the assumption of unitarity leads to
the appearance of a firewall near the event horizon, which is known as
\textit{firewall paradox} \cite{Almheiri:2012rt}, and this contradicts the
\textquotedblleft no-drama\textquotedblright\ principle in general relativity.

Island rule
\cite{Penington:2019npb,Almheiri:2019psf,Almheiri:2019hni,Almheiri:2019yqk,Almheiri:2020cfm,Engelhardt:2014gca}%
, as a way to resolve the information paradox, has been proposed in recent
years, and it can reproduce Page curve while avoiding the firewall problem at
the same time, attracting a lot of attention. The rule states that the entropy
of Hawking radiation is given by \cite{Almheiri:2020cfm}
\begin{equation}
S_{R}=Min_{X}\left\{  ext_{X}\left[  \frac{\mathcal{A}(X)}{4G_{N}}%
+S_{c}(\Sigma_{R}\cup\Sigma_{I})\right]  \right\}  ,
\end{equation}
where $X$ is quantum external surface (QES), namely the boundary of island
$\partial I$, and $\mathcal{A}(X)$ is the area of QES. $S_{c}(\Sigma_{R}%
\cup\Sigma_{I})$ is the entanglement entropy of matter in the regions formed
by both the exterior of cutoff surface and islands. The region between cutoff
surface and QES is denoted as $\Sigma_{X}$. The matter in the region
$\Sigma_{X}\cup\Sigma_{R}\cup\Sigma_{I}$ is assumed to be in a pure state,
then $S_{c}(\Sigma_{X})=S_{c}(\Sigma_{R}\cup\Sigma_{I})$ according to the
properties of entanglement entropy. In other words, the radiation entropy is
equal to that of black hole, given by
\cite{Ryu:2006bv,Hubeny:2007xt,Faulkner:2013ana,Engelhardt:2014gca}
\begin{equation}
S_{BH}=Min_{X}\left\{  ext_{X}\left[  \frac{\mathcal{A}(X)}{4G_{N}}%
+S_{c}(\Sigma_{X})\right]  \right\}  . \label{gen}%
\end{equation}
The part in square bracket is called general entropy, denoted by $S_{g}$. The
entanglement entropy of black holes is obtained by taking the extremum of the
general entropy first and then finding the minimum value.

The calculation of the exact entanglement entropy is very difficult. Aside
from the case where the entanglement entropy in two dimensional
Jackiw-Teitelboim gravity can be calculated exactly
\cite{Almheiri:2019yqk,Almheiri:2019qdq}, most computations of entanglement
entropy in black holes involve two approximation methods: the cutoff surface
is far from horizon \cite{Almheiri:2019psf}, and the cutoff surface is near
the horizon \cite{Casini:2009sr}, which had been applied in many different
kinds of black holes such as the eternal Schwarzschild black hole
\cite{Hashimoto:2020cas}, Reissner-Nordstr\"{o}m black hole
\cite{Wang:2021woy}, {lower-dimensional Gauss-Bonnet gravity
\cite{Anand:2022mla}, rotating Kerr black hole \cite{Wang:2024itz}, warped AdS
black hole \cite{Anand:2023ozw}}, and one-sided asymptotically flat black hole
\cite{Gan:2022jay,Guo:2023gfa}, among others. More generally, Ref.
\cite{Yu:2024fks} considered the island rule under a general metric function,
which not only maintains a finite radiation entropy but also ensures that the
island is always located outside horizon \cite{He:2021mst}. {And Ref.
\cite{Du:2022vvg} discusses the unitary constraints on the existence of
islands in $D\geq4$ Schwarzschild spacetime. }However, Ref.
\cite{Arefeva:2021kfx} argues that island configurations do not provide a
finite entanglement entropy when considering the decreasing black hole mass
due to evaporation. For the Schwarzschild black hole, the main reason is that
Hawking temperature diverges in the late stage of evaporation.

In fact, when a black hole evaporates close to the Planck scale, the
low-energy effective field theory approximation will no longer be applicable,
and new physics must be considered. These modifications, such as the
generalized uncertainty principle \cite{Maziashvili:2005pp}, regular black
holes \cite{frolov1990black,Hayward:2005gi}, loop quantum gravity
\cite{Modesto:2005zm,Hossenfelder:2009fc}, and the noncommutative geometry
\cite{Nicolini:2005vd,Nicolini:2008aj,Zhang:2017zy,Zhang:2024slz}, will
prevent further black hole evaporation, leading to a remnant. This paper
focuses on the island rule under noncommutative geometry.

When the spatial noncommutativity is considered, a completely different type
of black hole or noncommutative black hole (NBH) was obtained
\cite{Nicolini:2005vd,Myung:2006mz,Banerjee:2008du,Banerjee:2008gc}. The
Hawking temperature of NBH does not diverge as the mass decreases but
eventually approaches zero, which leads to the remnant. Whether the existence
of the remnant can lead to a consistent radiation process following the Page
curve is unclear. Thus, the requirement of the island for the NBH is the
purpose of our study in this paper. We will show that the island is necessary
for the recovery of the Page curve in the radiation process and discuss how
the island can help reproduce the Page curve for the NBH.

\section{Noncommutative black holes}

Start with the metric of the NBH \cite{Nicolini:2005vd},
\begin{equation}
ds^{2}=-f(r)dt^{2}+f^{-1}(r)dr^{2}+d\Omega^{2}, \label{nbhm}%
\end{equation}
where $f(r)=\frac{4M}{r\sqrt{\pi}}\gamma\left(  \frac{3}{2},\frac{r^{2}%
}{4\vartheta}\right)  $, $\gamma\left(  \frac{3}{2},\frac{r^{2}}{4\vartheta
}\right)  =\int_{0}^{r^{2}/4\vartheta}t^{1/2}e^{-t}dt$ is the lower incomplete
Gamma function, $\vartheta$ is the non-commutative parameter which is
considered to be a small (Plank length) positive number, and $\Omega$ is the
solid angle in this spacetime. A detailed derivation for NBHs refers to Ref.
\cite{Nicolini:2005vd}.

Different from the commutative Schwarzschild black hole (CBH), the NBH has two
horizons that is the inner horizon $r_{-}$ and outer horizon $r_{+}$,
respectively, given by $f(r)=0$. The outer horizon is often referred to as the
event horizon, $r_{h}\equiv r_{+}$. In particular, when the black hole mass
decreases to $M_{0}\simeq1.9\sqrt{\vartheta}$, the inner and outer event
horizons coincide. If the mass is below $M_{0}$, no black hole exists.

NBHs belong to spherically symmetric spacetime, and the extension can be
conducted using a standard method known as Kruskal extension
\cite{mtw1973,Arraut:2010qx}. For simplification, we will work in a fixed
angular coordinate system, setting $d\Omega^{2}=0$. Introduce the tortoise
coordinate $r_{\ast}=\int f^{-1}(r)dr$ and double null coordinates
$u=t-r_{\ast}$,\quad$v=t+r_{\ast}$, and then the metric (\ref{nbhm}) is
written as $ds^{2}=-f(r)dudv$. Further, define coordinate transformations,
$U=-e^{-\kappa u},V=e^{\kappa v}$ for the Right region, $U=e^{-\kappa
u},V=-e^{\kappa v}$ for the Left region, $U=e^{-\kappa u},V=e^{\kappa v}$ for
the Black hole region, and $U=-e^{-\kappa u},V=-e^{\kappa v}$ for the White
hole region. See the Penrose diagram for the details in the Supplemental
Material (SM). Thus the metric becomes
\begin{equation}
ds^{2}=-\Omega^{2}dUdV, \label{mem}%
\end{equation}
where $\Omega$ is the conformal factor given by $\Omega^{2}=\frac{|{f(r)|}%
}{\kappa^{2}}\exp[-2\kappa r_{\ast}]$. The spacetime structure of the NBH
resembles that of Reissner-Nordstr\"{o}m black hole due to the existence of
double horizons, as presented by the Penrose diagram in SM.

A dazzling feature for NBHs is that the Hawking temperature no longer diverges
as mass decreases, and it vanishes when the black hole mass reaches $M_{0}$.
As a result, the noncommutativity restricts evaporation process to a
Plank-size remnant, which is thermodynamically stable because of positive heat
capacity \cite{Myung:2006mz}. Since the noncommutativity works in the last
stage of evaporation, the evaporation time can still be estimated through the
Stefan-Boltzmann law, $\frac{dE}{dt}\approx-\sigma\mathcal{A}T^{4}$, where the
area of horizon $\mathcal{A}=4\pi r_{h}^{2}$ and Hawking temperature
$T=\kappa/2\pi$ both are function of the mass (or energy). Thus, the
evaporation time of a black hole, evolving from initial mass $M_{i}$ to finial
mass $M_{f}$, is given by
\begin{equation}
t=-\int_{M_{i}}^{M_{f}}\frac{1}{\sigma\mathcal{A}T^{4}}dM. \label{time}%
\end{equation}
where $\sigma=\pi^{2}/60$ is taken for a single massless field with two
degrees of freedom \cite{Bonanno:2006eu,Myung:2006mz} in the following calculation.

Eq. (\ref{time}) can be applied to both commutative and noncommutative black
holes, respectively. See the Fig. 3 in the SM for the presentation. While the
black hole evaporates completely in a finite amount of time in the commutative
case, a remnant with a long lifetime emerges in the noncommutative scenario.
The mass of remnant asymptotically approaches the extremal black hole mass
$M_{0}$ as the NBH evaporates but can never actually reach it
\cite{Chen:2015coy}.

\section{Hawking radiation with islands}

At first, the noncommutativity cannot restore the Page curve, which is seen by
the increasing radiation entropy with time, as calculated and presented in SM.
So the island is necessary even for the NBH. In this section, we will consider
the radiation process of the NBH with the contribution of islands. When the
island is present, it should be considered a part of radiation, as presented
in Fig. 1.

\begin{figure}[pth]
\centering
\includegraphics[width=0.6\linewidth]{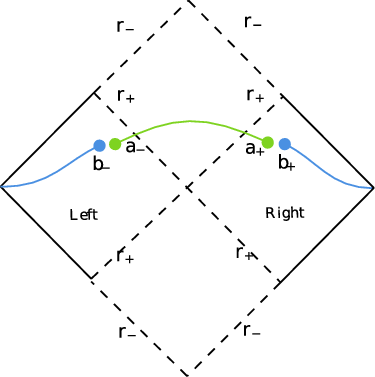} \caption{The schematic diagram
with islands. $a_{-}$ and $a_{+}$ represent the location of islands in the
Left and Right regions, respectively. $b_{-}$ and $b_{+}$ represent the
location of the cutoff surfaces in the Left and Right regions, respectively.}%
\label{Fig1}%
\end{figure}

\subsection{Cutoff surface far from horizon}

When the cutoff surface is far from horizon, the assumption of $s$-wave
approximation is effective. The entanglement entropy in $\Sigma_{X}$ is
represented as \cite{Almheiri:2019psf}
\begin{equation}
S_{c}=\frac{c}{3}\ln d(A,X), \label{sff}%
\end{equation}
where $d(A,X)$ is the distance between cutoff surface $A$ and QES $X$. In the
context of NBH, noncommutativity are confined to the spatial coordinate
components \cite{Nicolini:2005vd}. When focusing on the s-wave approximation,
we are concerned solely with the ($t,r$) components, in which case the
noncommutativity no longer play a role. Consequently, the form of Eq.
(\ref{sff}) remains valid. However, its expression involves the computation of
the spacetime distance $d(A,X)$, which is modified by the noncommutative black
hole metric. This introduces the influence of noncommutative parameters into
the entanglement entropy.

Considering a conformal transformation such as $ds^{2}=-\Omega^{2}dx^{+}%
dx^{-}$, the entropy can be rewritten as
\begin{equation}
S_{c}=\frac{c}{6}\ln\left[  \Omega(A)\Omega(X)\left(  x^{+}(A)-x^{+}%
(X)\right)  \left(  x^{-}(X)-x^{-}(A)\right)  \right]  . \label{sem}%
\end{equation}

Using Eq. (\ref{gen}) and (\ref{sem}), the general entropy for the case
presented in Fig. 1 is given by
\begin{equation}
S_{g}\ =\frac{2\pi a^{2}}{G_{N}}+\frac{c}{3}\ln\frac{d(a_{-},a_{+}%
)d(b_{-},b_{+})d(b_{-},a_{-})d(a_{+},b_{+})}{d(b_{-},a_{+})d(a_{-},b_{+})}.
\end{equation}
At the late time, we have
\begin{align}
d(a_{+},a_{-})  &  \simeq d(b_{+},b_{-})\simeq d(a_{+},b_{-})\nonumber\\
&  \simeq d(a_{-},b_{+})\gg d(a_{+},b_{+})\simeq d(a_{-},b_{-}).
\end{align}
Hence, the entropy can be simplified as
\begin{equation}
S_{g}\ \simeq\frac{2\pi a^{2}}{G_{N}}+\frac{c}{3}\ln[d(b_{-},a_{-}%
)d(a_{+},b_{+})]. \label{genf}%
\end{equation}

When the island is completely within the horizon, it is not hard to prove that
the derivative of the general entropy with regard to the time is
non-vanishing, which means that the extremum of the general entropy cannot be
obtained (see the calculation in SM). Thus the island should extend beyond the
horizon; namely, the QES is located outside the horizon. The general entropy
is given by
\begin{align}
S_{g}  &  =\frac{2\pi a^{2}}{G_{N}}+\frac{c}{6}\ln[\frac{4f(a)f(b)}{\kappa
^{4}}(\cosh[\kappa(r_{\ast}(a)-r_{\ast}(b))]\nonumber\\
&  -\cosh[\kappa(t_{a}-t_{b})])^{2}].
\end{align}
By solving $\frac{\partial S_{g}}{\partial t_{a}}=0$ and $\frac{\partial
S_{g}}{\partial a}=0$, and according to the island rule, the radiation entropy
is obtained (see the SM for the detailed calculation) as
\begin{align}
S_{R}  &  \simeq\frac{2\pi r_{h}^{2}}{G_{N}}+\frac{c}{6}\ln\left[
\frac{2r_{h}}{\kappa^{3}}e^{2\kappa r_{\ast}(b)}\right] \nonumber\\
&  +\left(  \frac{c^{3}G_{N}^{2}f^{\prime\prime}(r_{h})}{3456\pi^{2}\kappa
r_{h}^{3}}+\frac{c^{2}G_{N}}{36\pi^{2}r_{h}^{2}}\right)  e^{-2\kappa r_{\ast
}(b)}. \label{rei}%
\end{align}

When considering black hole evaporation, the radiation entropy (\ref{rei})
will change over time. The specific results can be seen in Fig. 2, with the
upper plot corresponding to CBH and the lower plot corresponding to NBH.
Although during the evaporation process, the commutative case initially
resembles Page curve, the radiation entropy inevitably diverges in the late
stage of evaporation. The reason is that the temperature diverges at the end
of evaporation, which means that the sureface gravity $\kappa$ approaches zero
then and the second term in Eq. (\ref{rei}) dominates, and thus, the
divergence appears as presented in the upper plot of Fig. 2. In contrast, with
the inclusion of non-commutative corrections, we successfully reproduce Page
curve, and the radiation entropy in the late stage of evaporation perfectly
coincides with the Bekenstein-Hawking entropy of the black hole, since the
temperature approaches zero in the final stage of evaporation for this case.

\begin{figure}[pth]
\centering
\includegraphics[width=0.8\columnwidth]{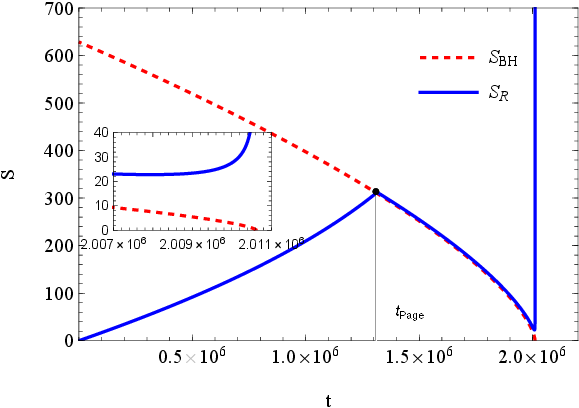}\newline%
\includegraphics[width=0.8\columnwidth]{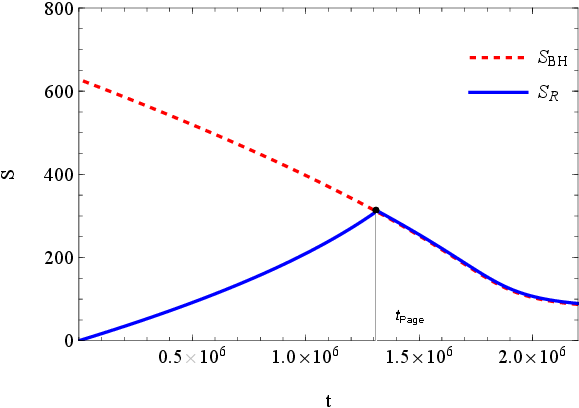}\newline.\caption{The upper and
lower plots correspond to CBH and NBH, respectively. The blue solid line
represents the radiation entropy given by island formula, while the red dashed
line represents the Bekenstein-Hawking entropy of the black hole. The island
appears at time $t_{Page}$. Considering the island, the radiation entropy of
CBH will exhibit a reduction, but it still inevitably diverges in the late
stages of evaporation, primarily due to the late-stage divergence of the
Hawking temperature. In contrast, for the NBH, the radiation entropy given by
island formula perfectly reproduces the Page curve. The inserted figure in the
upper one presents the behaviours of entropy near the time $t=2\times10^{6}$}%
\label{Fig2}%
\end{figure}

\subsection{Cutoff surface near horizon}

When cutoff surface is near horizon, the entanglement entropy is given by
\cite{Casini:2009sr}
\begin{equation}
S_{c}=-kc\frac{\mathcal{A}(A)}{L^{2}},
\end{equation}
where $\mathcal{A}(A)$ is the area of cutoff surface, and $L$ is the geodesic
distance between cutoff surface and QES. $c$ is the central charge of matter
fields, and $k$ is a constant relating to the matter field. The mathematical
form remains valid in the spatial noncommutativity, and the reason is the same
as interpreted for the entropy in Eq. (\ref{sff}). Therefore, for the NBH, the
general entropy is given by
\begin{align}
S_{g}  &  =\frac{2\pi a^{2}}{G_{N}}-2kc\frac{4\pi b^{2}}{L^{2}}\nonumber\\
&  =\frac{2\pi a^{2}}{G_{N}}-4kc\frac{\pi b^{2}\kappa}{\left(  \sqrt{b-r_{h}%
}-\sqrt{a-r_{h}}\right)  ^{2}},
\end{align}
where $L$ is the geodesic distance between $a$ and $b$, given as
$L=\int_{a_{+}}^{b_{+}}\sqrt{|{ds^{2}}|}=\int_{a}^{b}\frac{dr}{\sqrt{f(r)}%
}\simeq\frac{dr}{\sqrt{2\kappa(r-r_{h})}}=\sqrt{\frac{2}{\kappa}}\left(
\sqrt{b-r_{h}}-\sqrt{a-r_{h}}\right)  $.

After the extremum of the general entropy is taken (see SM for the detailed
calculation), the radiation entropy is obtained as
\begin{equation}
S_{R}=\frac{2\pi r_{h}^{2}}{G_{N}}-4\pi kcr_{h}^{2}\frac{\kappa}{b-r_{h}}.
\label{reh}%
\end{equation}

\begin{figure}[pth]
\centering
\includegraphics[width=0.8\columnwidth]{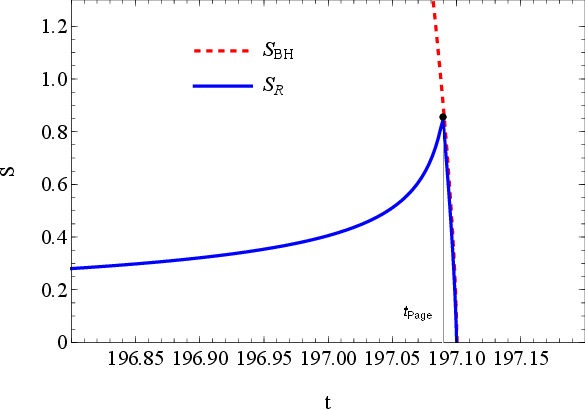} \newline%
\includegraphics[width=0.8\columnwidth]{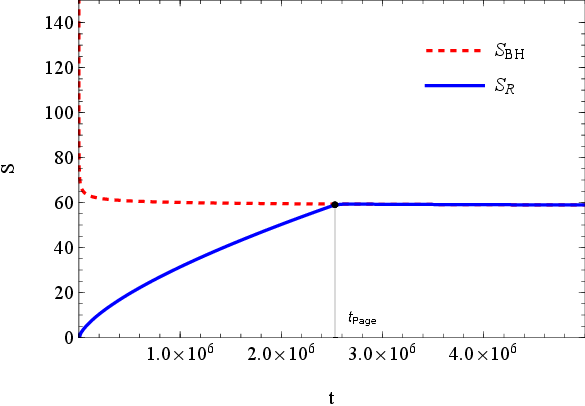} \newline\caption{The blue
solid line represents the radiation entropy, and the red dashed line
represents the black hole entropy. The upper and lower plots correspond to
CBHs and NBHs, respectively. Both reproduce the Page curve, but the Page time
for the NBH is greatly delayed.}%
\label{Fig3}%
\end{figure}

Since the cutoff surface is near the event horizon, the Hawking temperature
should be replaced by the local temperature $T_{loc}=T/\sqrt{f(r)}$ to
recalculate the evaporation time in Eq. (\ref{time}). Thus, the evaporation
time of the black hole is significantly shortened, due to the substantial
increase in the local Hawking temperature near the event horizon.

Now we consider the trend of radiation entropy as a function of time during
evaporation. In this case, the radiation entropy (\ref{reh}) will change with
time. In the early stages of evaporation without islands, we still assume the
validity of the $s$-wave approximation, and the final result is shown in Fig.
3. Note that the radiation entropy has a rapid decrease after the Page time as
seen in the upper plot of Fig. 3, which is caused by the rapid black hole
evaporation process from the perspective of a near-horizon observer since the
local Hawking temperature tends to increase rapidly after the Page time. It is
worth mentioning that, at this stage, both commutative and non-commutative
black holes can reproduce the Page curve. The Page time occurs in the late
stage of evaporation for the CBH, after which the contribution from islands
prevents the entropy from diverging. In contrast to CBH, the Page time for NBH
is greatly delayed (by four orders of magnitude), presenting the significant
role played by the spatial noncommutativity.

\section{Conclusion}

In this letter, we investigate the island rule under noncommutative geometry.
For evaporating black holes, the radiation entropy without islands will
diverge in both commutative and noncommutative black holes, so considering the
contribution of islands is significant for reproducing the Page curve. When
cutoff surface is far from event horizon, the existence of islands in the CBH
can temporarily reduce the radiation entropy, but eventually, due to the
divergent Hawking temperature, the entropy will diverge again, which is
unavoidable. After considering noncommutative effects, the situation improves
significantly, and the radiation entropy after the Page time perfectly matches
the black hole entropy. When cutoff surface is close to horizon, both
commutative and non-commutative black holes can reproduce the Page curve, but
the Page time for the non-commutative geometry is greatly delayed, allowing us
to avoid worrying about the contribution of islands for a long period of time.

In a word, noncommutative effects can improve the divergence of radiation, but
to strictly control its divergence, the contribution of islands is still
necessary. Moreover, based on the existing derivations, it can be expected
that any model with a zero-temperature remnant will improve the divergence of
radiation entropy in the late stage of evaporation. This also suggests a
potential avenue through which quantum corrections could influence the issue
of black hole information loss.

\bigskip

\acknowledgments

This work is supported by National Natural Science Foundation of China (NSFC)
with Grant No. 12375057 and the Fundamental Research Funds for the Central
Universities, China University of Geosciences (Wuhan).

\end{document}